# FlexPINN: Modeling Fluid Dynamics and Mass Transfer in 3D Micromixer Geometries Using a Flexible Physics-Informed Neural Network

Meraj Hassanzadeh[1], Ehsan Ghaderi[1], Mohamad Ali Bijarchi[*,1]

[1]Department of Mechanical Engineering, Sharif University of Technology, Tehran, Iran

## Abstract

In this study, fluid flow and concentration distribution inside a 3D T-shaped micromixer with various fin shapes and configurations are investigated using a Flexible Physics-Informed Neural Network (FlexPINN), which includes modifications over the vanilla PINN architecture. Three types of fins—rectangular, elliptical, and triangular—are considered to evaluate the influence of fin geometry, along with four different fin configurations inside the 3D channel to examine the effect of placement. The simulations are conducted at four Reynolds numbers: 5, 20, 40, and 80, in both single-unit (four fins) and double-unit (eight fins) configurations. The goal is to assess pressure drop coefficient, mixing index, and mixing efficiency using the FlexPINN method. Given the challenges in simulating 3D problems with standard PINN, several improvements are introduced. The governing equations are injected into the network as first-order, dimensionless derivatives to enhance accuracy. Transfer learning is used to reduce computational cost, and adaptive loss weighting is applied to improve convergence compared to the vanilla PINN approach. These modifications enable a consistent and flexible architecture that can be used across numerous tested cases. Using the proposed FlexPINN method, the pressure drop coefficient and mixing index are predicted with maximum errors of 3.25% and 2.86%, respectively, compared to Computational Fluid Dynamics (CFD) results. Among all the tested cases, the rectangular fin with configuration C in the double-unit setup at Reynolds number 40 shows the highest mixing efficiency, reaching a value of 1.63. The FlexPINN framework demonstrates strong capabilities in simulating fluid flow and species transport in complex 3D geometries.

## Introduction

Fluid mixing has consistently garnered significant attention in microfluidics research due to its critical role in enhancing the performance of microscale systems. This process is fundamental to a wide range of applications, including chemical synthesis, biomedical diagnostics, pharmaceutical analysis, and lab-on-a-chip systems [1–2]. A key challenge in micromixer design lies in the trade-off between mixing efficiency and pressure drop, enhancing one often compromises the other [3–4]. Generally, fluid mixing methods are categorized into two primary approaches: active techniques, which rely on external energy sources such as electric, magnetic, thermal, or acoustic fields to stimulate and improve mixing performance [5–10]; and passive techniques, which eliminate the need for external power by leveraging innovative channel design, increased interfacial contact between fluids, or the integration of structures like porous media and internal fins. In this study, we focus on the use of internal fins placed within a T-shaped microchannel as a passive strategy to improve mixing performance. This approach offers a simple yet effective enhancement without requiring additional equipment or operational costs [11–13].

---

[*] corresponding author: bijarchi@sharif.edu

Several studies have investigated the enhancement of mixing performance in microchannels through geometric modifications. Chen and Zhao [14] analyzed fluid flow in a 3D channel with internal obstacles to guide optimal obstacle design. They evaluated the impact of obstacle height, geometry, and number, concluding that height was the most influential parameter. Their results showed that using multiple obstacles improved mixing efficiency by up to 90%. Rasouli et al. [15] studied a curved micromixer enhanced with internal obstacles to promote normal advection and Dean vortices. Five geometric parameters — channel radius, angle, obstacle height and thickness, and aspect ratio — were optimized. At Reynolds numbers 3, 27, and 81, the design achieved a maximum mixing index error of 3% and a 7% pressure drop. Zou et al. [16] designed, fabricated, and evaluated an innovative 2D passive micromixer. The device achieved near-complete mixing at Re = 10 and showed improvements of 20%, 35%, and 15% at Re = 0.5, 1.25, and 10, respectively. Fabrication precision and design reliability were also significantly enhanced. Chen and Lv [17] selected three geometric ratios as design variables and considered outlet mixing index and pressure drop as dual objectives for optimization. Latin Hypercube Sampling (LHS) was used to sample the design space uniformly. At Re = 1 and Re = 10, mixing efficiency improved by 20.59% and 14.07% over the reference design. Ouro-Koura et al. [18] proposed a method to significantly enhance passive mixing at low Reynolds numbers (~10). By hydrophobic slip patterns on channel floors, high mixing indices were reported even at low Res. Agarwal and Wang [19] introduced non-aligned inlets combined with rigid dispersion plates and cylindrical obstacles. At Re ≈ 30, a mixing efficiency of 85% was achieved with a pressure drop of 2300 Pa. Their design, though simple, demonstrated effective performance in specific Reynolds regimes. Arrangements such as linear and stepped layouts, upward and downward flows, and the presence or absence of obstacles were compared to identify optimal setups. Najafpour et al. [20] numerically investigated flow in a 3D T-shaped channel with twisted geometry. They assessed the influence of different twisting patterns on the mixing index across various Reynolds numbers. Depending on the twist type, the mixing index improved by up to 80%. Cunegatto et al. [21] introduced high-performance passive micromixer designs featuring multiple obstacles. These micromixers used Y-shaped inlets with circular barriers in repeated cells along the device. Using CFD simulations, the effects of vertical and horizontal spacing between obstacles on mixing percentage, pressure drop, and Mixing Energy Cost (MEC) were analyzed. Vertical spacing had a greater influence on mixing, while both directions significantly affected pressure drop. Barzoki [22] conducted a numerical study using the finite element method to evaluate fluid flow and mass transfer characteristics in novel micromixers featuring column arrays. Using 2D geometries, the study assessed mixing performance based on concentration distribution and mixing index. The combination of fast mixing, low-pressure drop, and short mixing length made these designs highly promising for microfluidic applications.

Conventional numerical simulations in fluid dynamics are predominantly based on spatial and/or temporal discretization of the governing equations using polynomial approximations, which are then transformed into a finite-dimensional algebraic system. However, due to the multiscale nature of physical phenomena and the high sensitivity of solutions to mesh quality—especially in complex geometries—these methods can become prohibitively expensive and time-consuming, particularly for real-time applications, optimal design, or uncertainty quantification [23–25]. In recent years, deep learning techniques have emerged as an alternative or complementary tool alongside traditional numerical methods for fluid flow modeling. Several studies have explored the use of artificial neural networks trained on data obtained from experiments or CFD simulations to infer critical flow parameters [26–29]. Despite promising results, conventional machine learning approaches are often highly data-dependent, which limits their applicability in many engineering contexts where large-scale datasets are either unavailable or infeasible to generate [30–31]. Moreover, many engineering problems are governed by well-understood physical laws, typically expressed as differential equations—information that remains unused in purely data-driven models [32]. To bridge this gap, physics-informed neural networks (PINNs) were introduced [33]. These models incorporate the governing differential equations of the physical system, along with any initial and boundary conditions, directly into the training process. As a result, the network learns not only from available data but also from the underlying physics of the problem, enabling accurate predictions with minimal reliance on large datasets [34]. Researchers have applied the PINN methodology across a wide range of fluid flow problems. For instance, Sun et al. [35] explored three cardiovascular flow scenarios: flow in a circular pipe, flow through a stenosed (narrowed) artery, and flow in an aneurysmal (dilated) vessel. Their results demonstrated a high degree of agreement between physics-constrained neural network models and conventional CFD simulations. The study also examined the impact of advanced adaptive activation functions, comparing the performance of the proposed label-free PINN framework to traditional data-driven models in terms of accuracy and efficiency. In another work, Cai et al. [36] used a PINN approach to reconstruct continuous 3D velocity and pressure fields from time-resolved 3D temperature measurements obtained via Tomographic Background-Oriented Schlieren (Tomo-BOS) imaging. By integrating governing physical laws with limited experimental data, they successfully extracted hidden quantities of interest using only sparse observational input. Chen et al. [37] focused on discovering governing Partial Differential Equations (PDEs) from limited and noisy datasets. Their results, validated through both numerical and experimental means, confirmed the robustness and accuracy of the PINN framework for identifying a wide range of PDE systems, even under severe data scarcity, noise, and varying initial/boundary conditions. In a separate study, Bararnia and Esmaeilpour [38] analyzed three benchmark problems—the Blasius-Pohlhausen boundary layer, the Falkner-Skan equation, and natural convection-driven flows—to investigate how the nonlinearity of governing equations and unbounded boundary conditions affect the neural network structure, particularly in terms of network width (number of neurons per layer) and depth (number of layers). Their trained models were successfully applied to previously unseen data to accurately estimate boundary layer thickness. Yang et al. [39] explored various adaptive sampling methods for collocation points to assess their effectiveness within the physics-informed neural network (PINN) framework. The overall performance of the data-driven PINN (DD-PINN) framework, capable of leveraging data obtained from digital twin scenarios, was then evaluated. The scalability of this approach for more general physics was confirmed within the context of the Navier-Stokes equations, where PINNs can provide accurate responses at different Reynolds numbers without the need for retraining. In addition to fluid flow, PINN-based studies have been extended to heat transfer [40-42], turbulence modeling [43-45], and biofluid mechanics [46-48].

In the realm of mass transfer and microfluidic lab-on-chip applications [49], two PINN architectures—single-network and segregated-network—were employed to predict momentum, chemical species, and temperature distributions in the problem of dry air humidification in a simple two-dimensional rectangular domain. Both architectures were trained with varying hyperparameter configurations (such as network width and depth) to identify the optimal setup for enhanced performance. While the single-network model struggled with maintaining species conservation in different regions of the computational domain, the segregated-network model effectively adhered to the species conservation law. In another study, Sun et al. [50] compared and validated PINN model predictions, including fluid flow, electric potential distribution, and ion transport characteristics, against results obtained from Finite Element Methods (FEM). Overall, the PINN model demonstrates significant promise for achieving accurate solutions in microfluidic systems with multiphysics coupling. Furthermore, in the context of micromixers, Chang et al. [51] explored an electro-osmotic active micromixer. Their study predicted the flow parameters and electric potential based on solute concentration. In regions near the electrodes, the flow velocity increased up to three orders of magnitude higher than the average, leading to considerable errors in the PINN model's velocity predictions. In summary, these results highlight the advantages of the proposed approach and present a promising future for PINN-based methods in multiphysics predictions for other microfluidic devices. A review of previous studies shows that so far, no research has been conducted on the use of the PINN method in 3D micromixers. On the other hand, due to the inherently three-dimensional nature of the mixing phenomenon, a 3D analysis is essential. However, because solving 3D problems using the PINN method is quite challenging, only a few studies — such as those by Zhang and Zhao [52] and Biswas and Anand [53] — have investigated flow in simple channels without the presence of fins.

In this study, an innovative method called FlexPINN is proposed to address the challenges encountered in simulating fluid mixing within three-dimensional microfluidic channels featuring obstacles. These channels contain obstacles with various shapes, such as rectangular, triangular, and elliptical fins, which complicate the flow behavior. The FlexPINN method builds upon the traditional physics-informed neural network (PINN) architecture and incorporates several enhancements, including the use of parallel and series neural networks, modifications to the governing equations (dimensionless first-order derivatives), and the application of adaptive learning techniques to improve convergence. Additionally, transfer learning is employed for network weight initialization, leading to a reduction in model computation time. This method represents a significant advancement in simulating fluid dynamics within complex systems compared to the vanilla PINN approach. The first section presents the physics of the problem, followed by an introduction to the FlexPINN method. In the results section, contour plots and graphs related to the fluid flow and mass transfer phenomena are provided.

## Problem Description

This study investigates the fluid flow within a T-shaped micromixer with varying fin geometries and arrangements. A schematic of the problem's geometry, along with its dimensions, is shown in Figure 1. Three different types of fins—rectangular, elliptical, and triangular—are considered in this research. The fin geometry is designed such that, for all three cases, the height (h) and width (w) of the fins are identical. Additionally, the study examines the channel with both a single unit and two units, as illustrated in the figure. Table 1 presents the dimensions and specifications of the three-dimensional channel, including the fins.

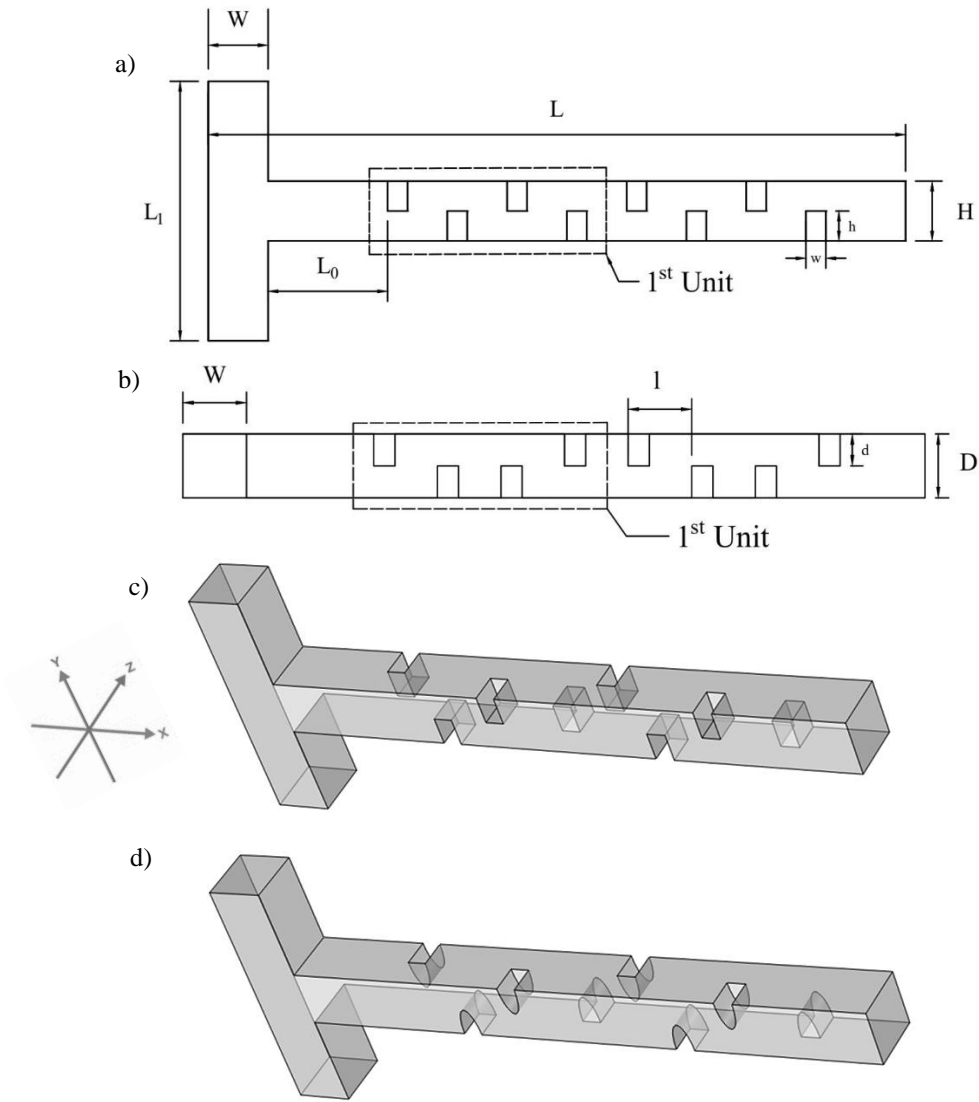

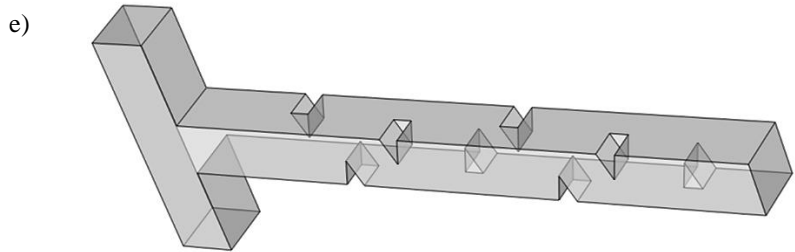

Figure 1: Schematic of the problem geometry: a) Channel dimensions with fins in the xy plane, b) Channel dimensions with fins in the xz plane, c) Rectangular fins, d) Elliptical fins, e) Triangular fins.

Table 1: Geometric specifications of the channel including fins

| | |
|---|---|
| L (1 Unit) | 2.4 mm |
| L (2 Units) | 3.5 mm |
| $L_0$ | 0.6 mm |
| $L_1$ | 1.3 mm |
| D | 0.3 mm |
| H | 0.3 mm |
| W | 0.3 mm |
| d | 0.15 mm |
| h | 0.15 mm |
| w | 0.1 mm |
| l | 0.3 mm |

Given that this study is a three-dimensional investigation, the different fin arrangements within the channel are also considered. In Figure 2, the four fin arrangements analyzed in the case of the channel with a single unit are presented.

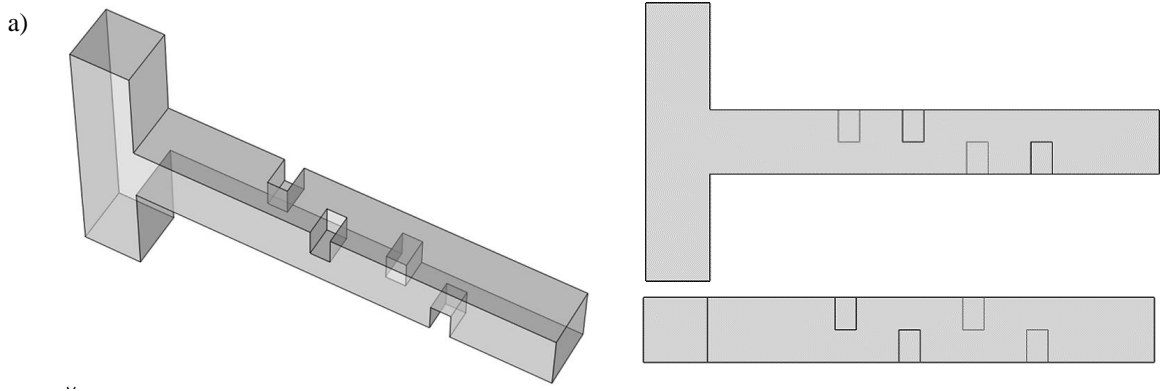

a)

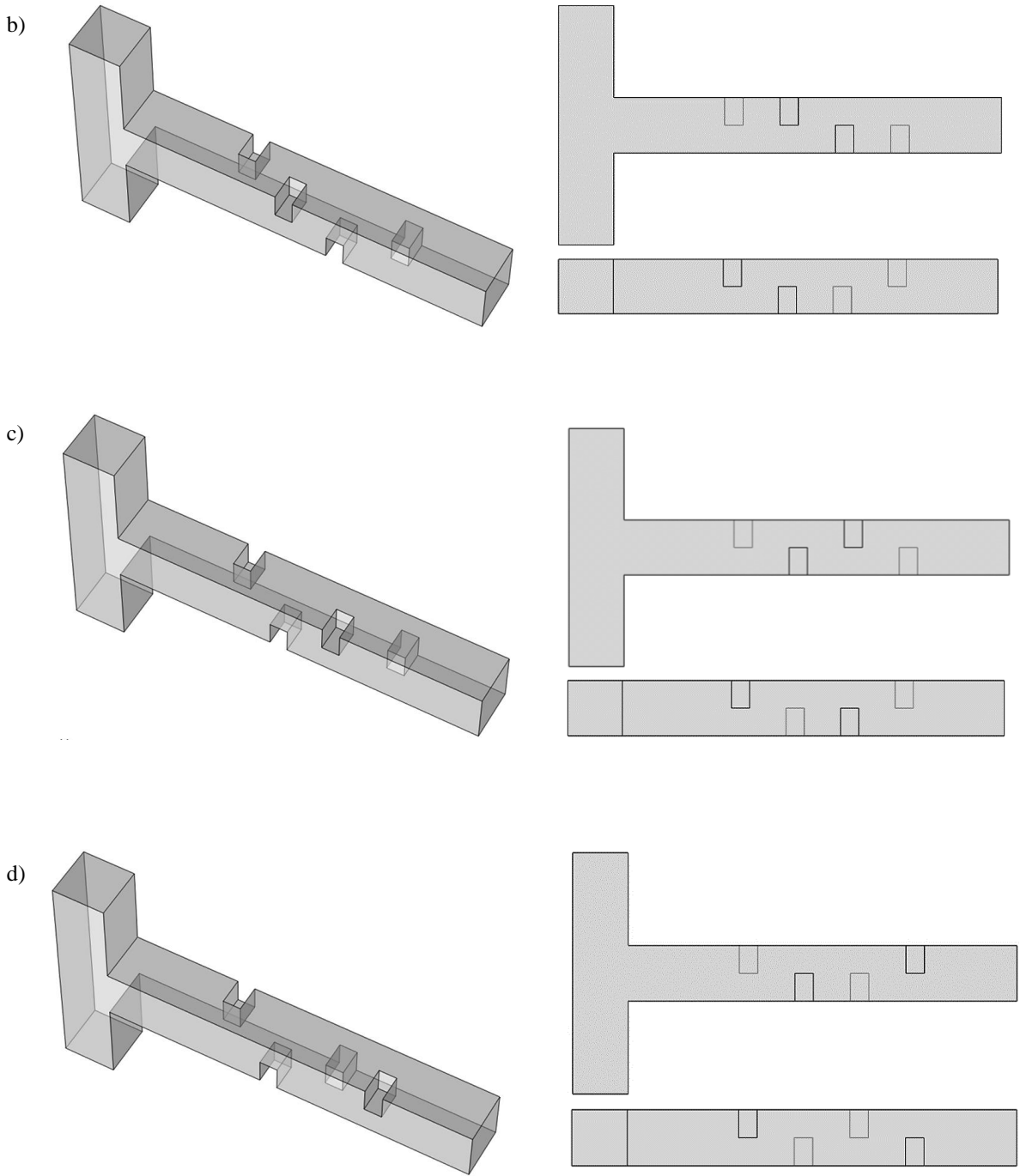

Figure 2: Different fin orientations in the channel with a single unit: a) Configuration A, b) Configuration B, c) Configuration C, d) Configuration D.

Assuming the continuity condition is satisfied in the dimensions of the micromixer, the governing differential equations of the problem in the Cartesian coordinates of the channel in steady-state three dimensional, are derived according to equations (1-3) [54]. These equations, which include the continuity, momentum, and mass transfer equations, are expressed in terms of lower-order and dimensionless derivatives in this study, according to equations (4-19). Equation (4) represents the continuity equation, equations (5-7) represent the momentum equations in three directions, and equation (14) represents the mass transfer equation. Other equations are constitutive equations that assist in the conservation equations. To expedite the solution process, these equations are written in the form of first-order derivatives [55], which ensures that the backward differentiation process (Backpropagation) occurs only once in the neural network. Moreover, to increase solution accuracy, the dimensionless form of the equations is used [49], which balances the terms in the loss function. In these relations, U is the velocity vector, u*, v*, and w* are the dimensionless velocity components in the x, y, and z directions, correspondingly. Additionally, p*, τ*, and c* represent the dimensionless pressure, dimensionless stress, and dimensionless concentration, respectively. Moreover, ρ, μ, $D_c$, and $U_m$ are the density, viscosity, diffusion coefficient of concentration, and the mean velocity of the fluid in the main channel, respectively. In these relations, J* is the dimensionless concentration flux, and the Reynolds and Schmidt numbers are defined in equations (15) and (16) [54].

$$\nabla \cdot \vec{U} = 0 \tag{1}$$

$$\rho(\vec{U} \cdot \nabla)\vec{U} = -\nabla p + \mu \nabla^2 \vec{U} \tag{2}$$

$$(\vec{U} \cdot \nabla)c = D_c \nabla^2 c \tag{3}$$

$$\frac{\partial u^*}{\partial x^*} + \frac{\partial v^*}{\partial y^*} + \frac{\partial w^*}{\partial z^*} = 0 \tag{4}$$

$$\left(u^* \frac{\partial u^*}{\partial x^*} + v^* \frac{\partial u^*}{\partial y^*} + w^* \frac{\partial u^*}{\partial z^*}\right) - \frac{\partial \tau_{xx}^*}{\partial x^*} - \frac{\partial \tau_{xy}^*}{\partial x^*} - \frac{\partial \tau_{xz}^*}{\partial x^*} = 0 \tag{5}$$

$$\left(u^* \frac{\partial v^*}{\partial x^*} + v^* \frac{\partial v^*}{\partial y^*} + w^* \frac{\partial v^*}{\partial z^*}\right) - \frac{\partial \tau_{xy}^*}{\partial y^*} - \frac{\partial \tau_{yy}^*}{\partial y^*} - \frac{\partial \tau_{yz}^*}{\partial y^*} = 0 \tag{6}$$

$$\left(u^* \frac{\partial w^*}{\partial x^*} + v^* \frac{\partial w^*}{\partial y^*} + w^* \frac{\partial w^*}{\partial z^*}\right) - \frac{\partial \tau_{xz}^*}{\partial z^*} - \frac{\partial \tau_{yz}^*}{\partial z^*} - \frac{\partial \tau_{zz}^*}{\partial z^*} = 0 \tag{7}$$

$$-p^* + \frac{2}{Re} \frac{\partial u^*}{\partial x^*} - \tau_{xx}^* = 0 \tag{8}$$

$$-p^* + \frac{2}{Re} \frac{\partial v^*}{\partial y^*} - \tau_{yy}^* = 0 \tag{9}$$

$$-p^* + \frac{2}{Re} \frac{\partial w^*}{\partial z^*} - \tau_{zz}^* = 0 \tag{10}$$

$$\frac{1}{Re}\left(\frac{\partial u^*}{\partial y^*} + \frac{\partial v^*}{\partial x^*}\right) - \tau_{xy}^* = 0 \tag{11}$$

$$\frac{1}{Re}\left(\frac{\partial u^*}{\partial z^*} + \frac{\partial w^*}{\partial x^*}\right) - \tau_{xz}^* = 0 \tag{12}$$

$$\frac{1}{Re}\left(\frac{\partial v^*}{\partial z^*} + \frac{\partial w^*}{\partial y^*}\right) - \tau_{yz}^* = 0 \tag{13}$$

$$\left(u^* \frac{\partial c^*}{\partial x^*} + v^* \frac{\partial c^*}{\partial y^*} + w^* \frac{\partial c^*}{\partial z^*}\right) + \frac{1}{ReSc}\left(\frac{\partial J_x^*}{\partial x^*} + \frac{\partial J_y^*}{\partial y^*} + \frac{\partial J_z^*}{\partial z^*}\right) = 0 \tag{14}$$

$$J_x^* + \frac{\partial c^*}{\partial x^*} = 0 \tag{15}$$

$$J_y^* + \frac{\partial c^*}{\partial y^*} = 0 \tag{16}$$

$$J_z^* + \frac{\partial c^*}{\partial z^*} = 0 \tag{17}$$

$$Re = \frac{\rho U_m D}{\mu} \tag{18}$$

$$Sc = \frac{\mu}{\rho D_c} \tag{19}$$

The boundary conditions used are given in equations (20-35), where equations (20-27) correspond to the inlet boundary conditions, equations (28-29) correspond to the outlet boundary conditions, and equations (30-35) represent the wall boundary conditions. At the inlet of the channel, parabolic velocity profiles are considered at both inlets, with the concentration boundary condition set to unity at the top and zero at the bottom of the T-shape. Additionally, at the outlet of the channel, the pressure and concentration flux boundary conditions are set to zero, while on the walls of the channel and fins, a wall boundary condition is applied (all velocity components and concentration flux are set to zero). In these relations, the subscript w refers to the wall, and n denotes the direction perpendicular to the boundary.

$$u^*\left(x^* = \left[0, \frac{W^*}{2}\right], y^* = \frac{L_1^* + D^*}{2}, z^* = \left[0, \frac{H^*}{2}\right]\right) = 0 \tag{20}$$

$$u^*\left(x^* = \left[0, \frac{W^*}{2}\right], y^* = \frac{-L_1^*}{2}, z^* = \left[0, \frac{H^*}{2}\right]\right) = 0 \tag{21}$$

$$v^*\left(x^* = \left[0, \frac{W^*}{2}\right], y^* = \frac{L_1^* + D^*}{2}, z^* = \left[0, \frac{H^*}{2}\right]\right) = -\left(\frac{16}{D^{*4}}\right)(z^*D^* - z^{*2})(x^*D^* - x^{*2}) \tag{22}$$

$$v^*\left(x^* = \left[0, \frac{W^*}{2}\right], y^* = \frac{-L_1^*}{2}, z^* = \left[0, \frac{H^*}{2}\right]\right) = \left(\frac{16}{D^{*4}}\right)(z^*D^* - z^{*2})(x^*D^* - x^{*2}) \tag{23}$$

$$w^*\left(x^* = \left[0, \frac{W^*}{2}\right], y^* = \frac{L_1^* + D^*}{2}, z^* = \left[0, \frac{H^*}{2}\right]\right) = 0 \tag{24}$$

$$w^*\left(x^* = \left[0, \frac{W^*}{2}\right], y^* = \frac{-L_1^*}{2}, z^* = \left[0, \frac{H^*}{2}\right]\right) = 0 \tag{25}$$

$$c^*\left(x^* = \left[0, \frac{W^*}{2}\right], y^* = \frac{L_1^* + D^*}{2}, z^* = \left[0, \frac{H^*}{2}\right]\right) = 1 \tag{26}$$

$$c^*\left(x^* = \left[0, \frac{W^*}{2}\right], y^* = \frac{-L_1^*}{2}, z^* = \left[0, \frac{H^*}{2}\right]\right) = 0 \tag{27}$$

$$p^*(x^* = L^*, y^* = [0, D^*], z^* = [0, W^*]) = 0 \tag{28}$$

$$J_x^*(x^* = L^*, y^* = [0, D^*], z^* = [0, W^*]) = 0 \tag{29}$$

$$u^*(x_w^*, y_w^*, z_w^*) = 0 \tag{30}$$

$$v^*(x_w^*, y_w^*, z_w^*) = 0 \tag{31}$$

$$w^*(x_w^*, y_w^*, z_w^*) = 0 \tag{32}$$

$$J_x^*(x_w^*, y_w^*, z_w^*) \cdot n_{x^*} = 0 \tag{33}$$

$$J_y^*(x_w^*, y_w^*, z_w^*) \cdot n_{y^*} = 0 \tag{34}$$

$$J_z^*(x_w^*, y_w^*, z_w^*) \cdot n_{z^*} = 0 \tag{35}$$

## Solution Methodology

In the present study, an enhanced Physics-Informed Neural Network (PINN) approach is employed to solve the governing equations, which include the continuity equation, momentum equations in the x, y, and z directions, and the mass transfer equation, as defined in Equations (4–17). A schematic of the network architecture is presented in Figure 3. The input to the network consists of the three spatial coordinates (x, y, z), which are fed into 14 parallel sub networks responsible for predicting the 14 desired output quantities. To accelerate the solution process, the equations are reformulated in terms of first-order derivatives using a variable substitution technique. This approach enables the backpropagation to compute derivatives only once, significantly reducing the computational cost. Furthermore, to maintain a balanced loss function and enhance the stability of the learning process, the non-dimensionalized form of the governing equations with first-order derivatives is used as input to the network.

Given the three-dimensional complexity of the problem and the associated challenges of applying conventional PINN methods in such settings, mass flux constraints at selected channel cross-sections are introduced as penalty terms in the loss function. Prior studies have recommended the inclusion of such penalty terms in problems with complex geometries or physics [56]. Additionally, the loss function weights are adaptively updated during training to further improve convergence. To accelerate the learning process, transfer learning is implemented: the network trained on rectangular fin geometry is initialized using Xavier initialization [57], and the final parameters of this network are then used to initialize networks for elliptical and triangular fin shapes. Further details regarding this implementation are provided in the results section.

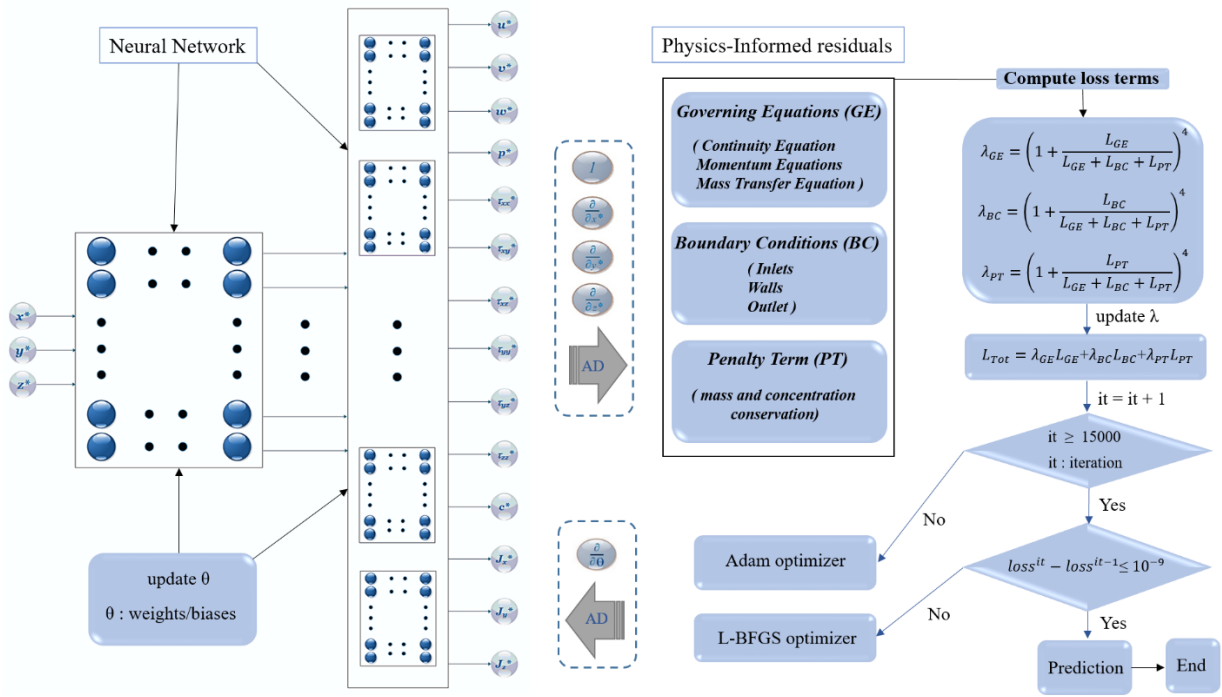

Figure 3: The FlexPINN architecture employed in this study.

To simulate the problem using the proposed method, random points were distributed using Latin Hypercube Sampling (LHS) [58] throughout the computational domain and along the boundaries of the geometry. Figure 4 shows the point distribution and their count on a representative xz-plane located at the mid-height of the main channel. The figure also indicates the planes where a constant mass flow rate was imposed as a penalty term in the loss function to improve convergence. For further details on the point distribution and their arrangement across other planes, please refer to the GitHub repository associated with this study.

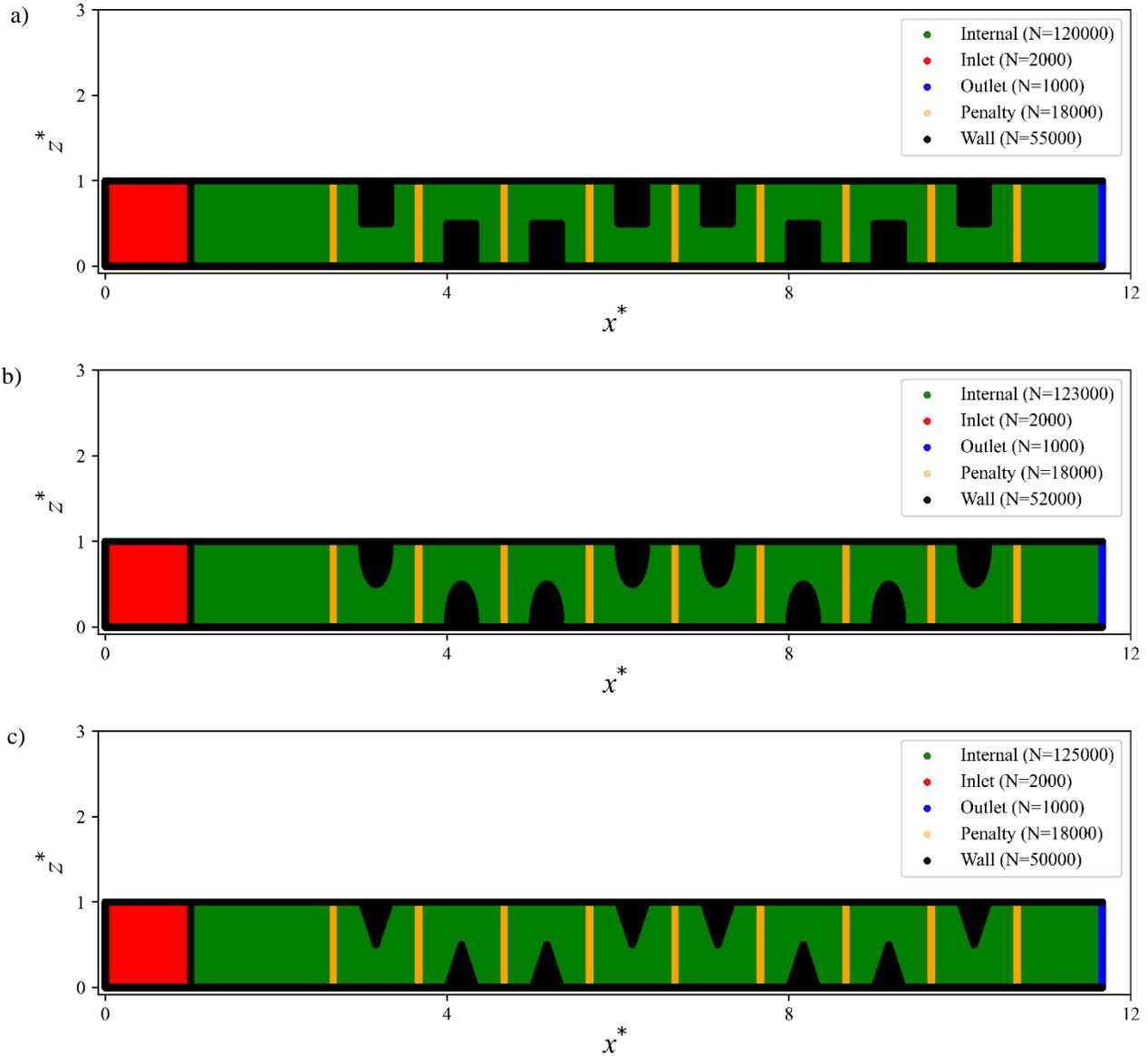

Figure 4: Point distributions generated using Latin Hypercube Sampling (LHS) for three different fin shapes, shown on an xz-plane at mid-height of the main channel. a) Rectangular fins, b) Elliptical fins, c) Triangular fins. (**Note:** The yellow slices indicate regions where a constant mass flow rate was enforced as a penalty term in the FlexPINN method).

## Results and Discussion

In this study, the FlexPINN method—whose network structure is illustrated in Figure 3—was employed to simulate the 3D fluid flow and mass transfer phenomena in a channel equipped with fins. The simulations were conducted for single-unit and double-unit channel, four different fin placement configurations, three fin shapes, and four Reynolds numbers (5, 20, 40, and 80) at a constant Schmidt number of 1000. At low Reynolds numbers (below ~100) where the flow remains laminar, mixing between two fluids in a straight channel is challenging due to the absence of turbulence, as molecular diffusion becomes dominant and decelerates the mixing mechanism. To address this, various fin configurations were employed to enhance mixing.

Figure 5 illustrates the reduction of the loss function terms for the rectangular fin shapes and the accelerated learning achieved using transfer learning for elliptical and triangular fins in a double-unit channel at Re = 5. The transfer learning approach used in FlexPINN resulted in a 35% reduction in training time for the elliptical and triangular fins. The sudden change in the loss curves at iteration 15,000 corresponds to the switch from the ADAM optimizer [59] to L-BFGS [60], implemented to improve solution accuracy. The implementation was carried out in Python using the PyTorch library, which supports Automatic Differentiation (AD) [61]. Simulations were performed on an RTX 4080 GPU with 16 GB RAM, requiring approximately 5.5 hours for rectangular fins and 4 hours for elliptical and triangular fins with transfer learning. Compared to the vanilla PINN approach, which takes around 8 hours per case, the proposed method significantly reduces computation time.

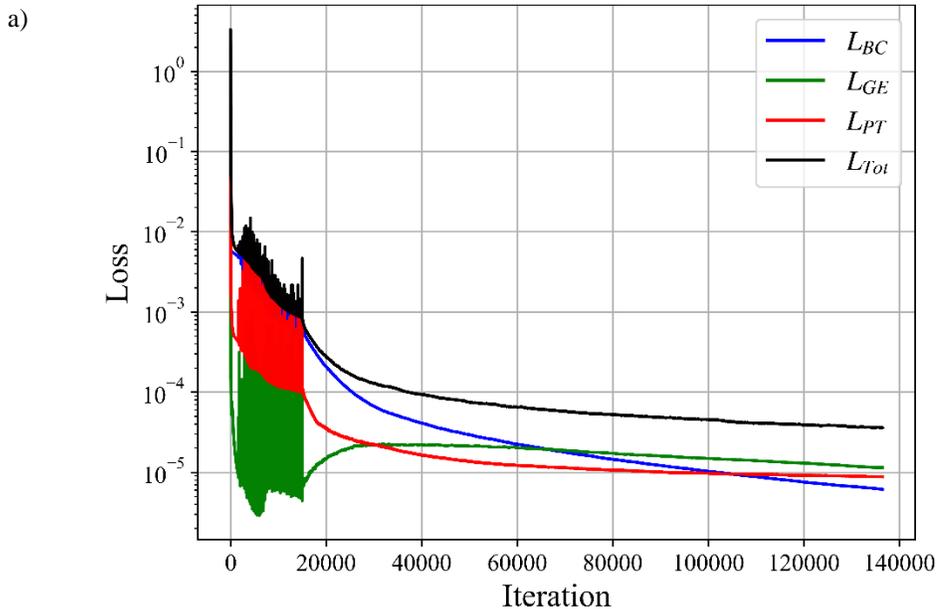

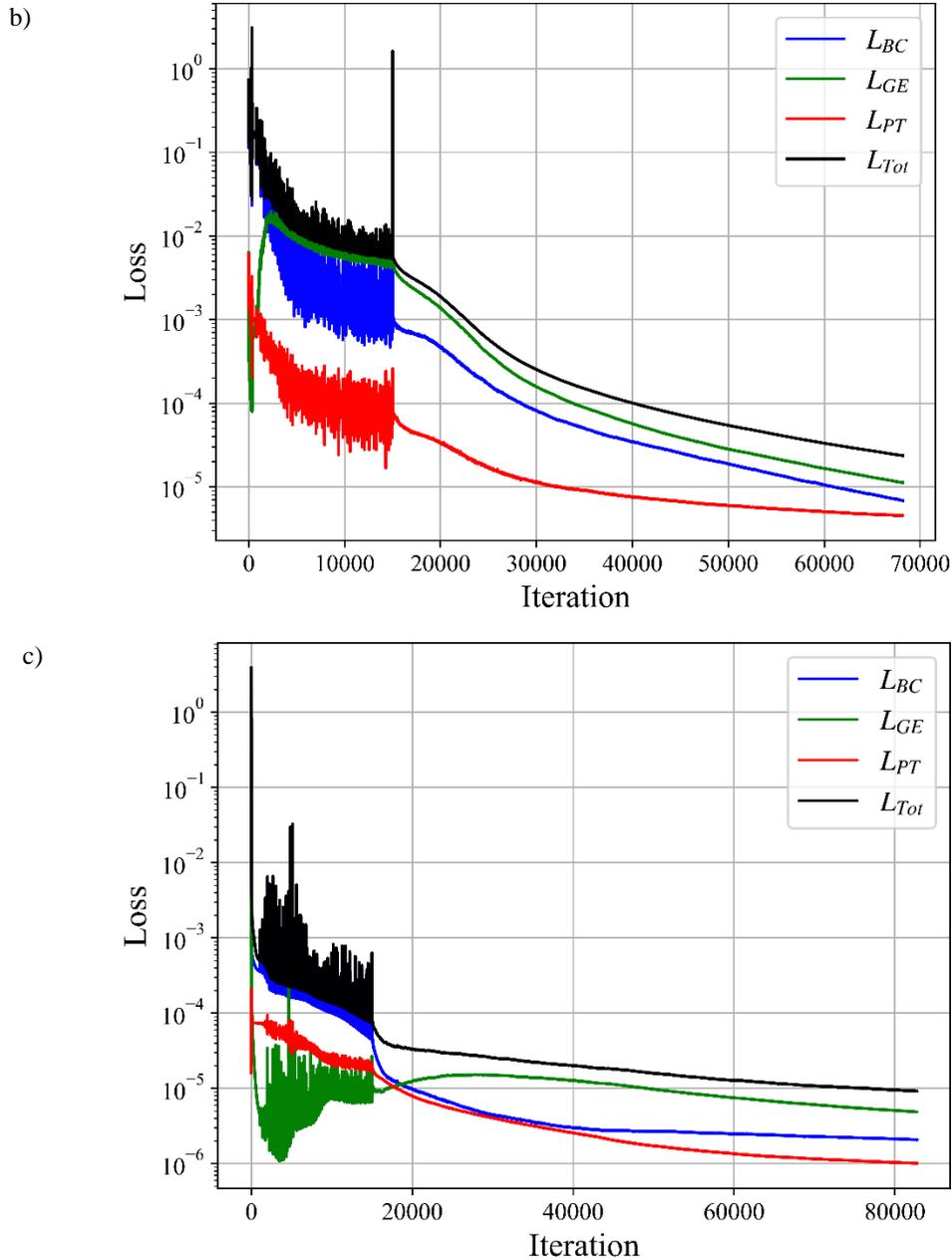

Figure 5: Loss function curves for boundary conditions, governing equations, penalty terms, and total loss. (a) Rectangular fins, (b) Elliptical fins, (c) Triangular fins.

Figure 6 compares the dimensionless concentration contours obtained using the vanilla PINN and the improved FlexPINN method for two different Reynolds numbers in the two-unit channel. As shown in the figure, due to the three-dimensional nature of the problem and the presence of fins inside the channel, the vanilla PINN fails to accurately predict the concentration distribution. In contrast, the improved FlexPINN method provides an accurate representation of the governing fluid dynamics.

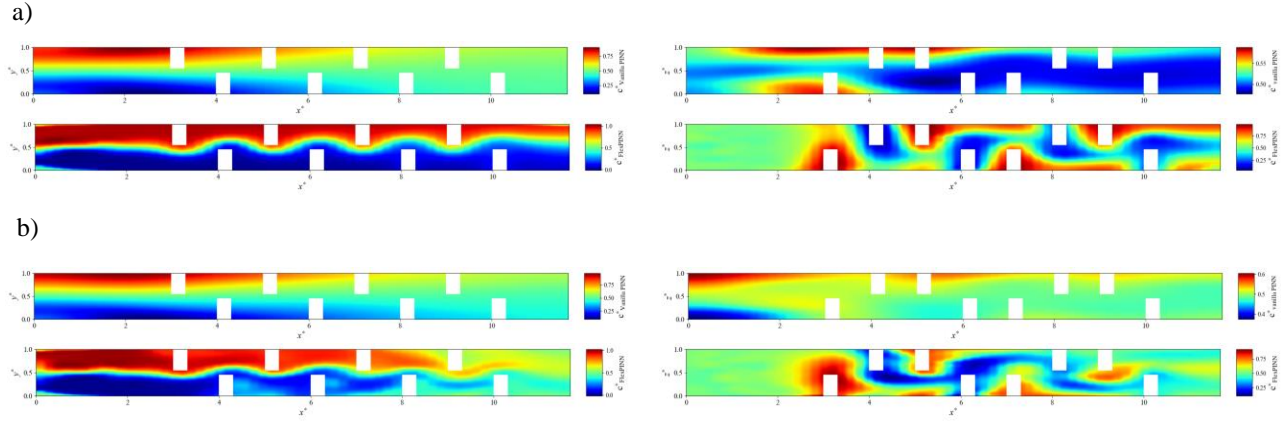

Figure 6: Comparison of dimensionless concentration contours between the FlexPINN and vanilla PINN methods in the xy-plane (left) and xz-plane (right).
(a) Re = 5   (b) Re = 80

## Validation

To verify the results obtained from the FlexPINN method, a comparison is made with the results of similar studies and other numerical methods in this section. As shown in Figure 7, the T-shape micromixer without fins is compared with the results from Al-Zoubi et al. [62]. Two metrics, the mixing index (Equation 37) and pressure drop coefficient (Equation 36), are used for comparison [17]. As shown in the figure, at low Reynolds numbers, the FlexPINN results slightly deviate from the results of Al-Zoubi et al. However, the highest error with respect to their study is reported as 2.86% in the mixing index and 3.25% in the pressure drop coefficient.

$$C_p = \frac{\Delta P}{\frac{1}{2}\rho U_m^2} \tag{36}$$

$$MI = 1 - \sqrt{\frac{1}{N}\sum_{i=1}^{N}\left(\frac{c_i^* - c_{mean}^*}{c_{mean}^*}\right)^2} \tag{37}$$

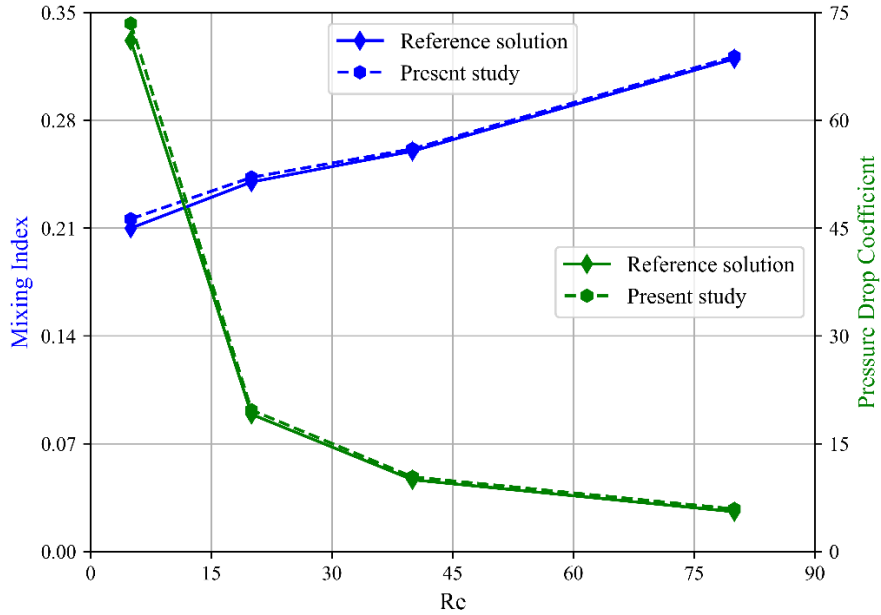

Figure 7: Comparison of FlexPINN results with the results obtained in the study by Al-Zoubi et al. [60] in the 3D channel without fins.

To further validation of the results obtained from the FlexPINN method, the results in the single unit case at Re = 5 were compared with those obtained from CFD solutions. In Figure 8, the contour plots of the dimensionless velocity components, concentration, and pressure in the channel with rectangular fins in configuration A at different sections along the channel are shown. As indicated in the figure, the results obtained from the FlexPINN method show high accuracy, with errors in the calculation of the dimensionless horizontal velocity in insignificant regions of the channel reported to be a maximum of 8%. Additional results for other configurations can be found at the [GitHub link](#).

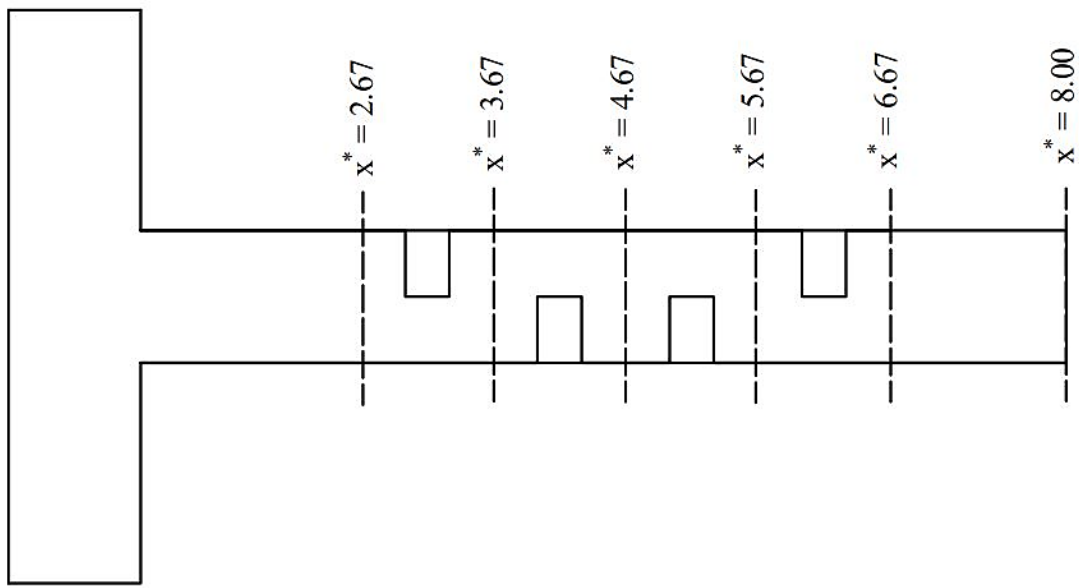

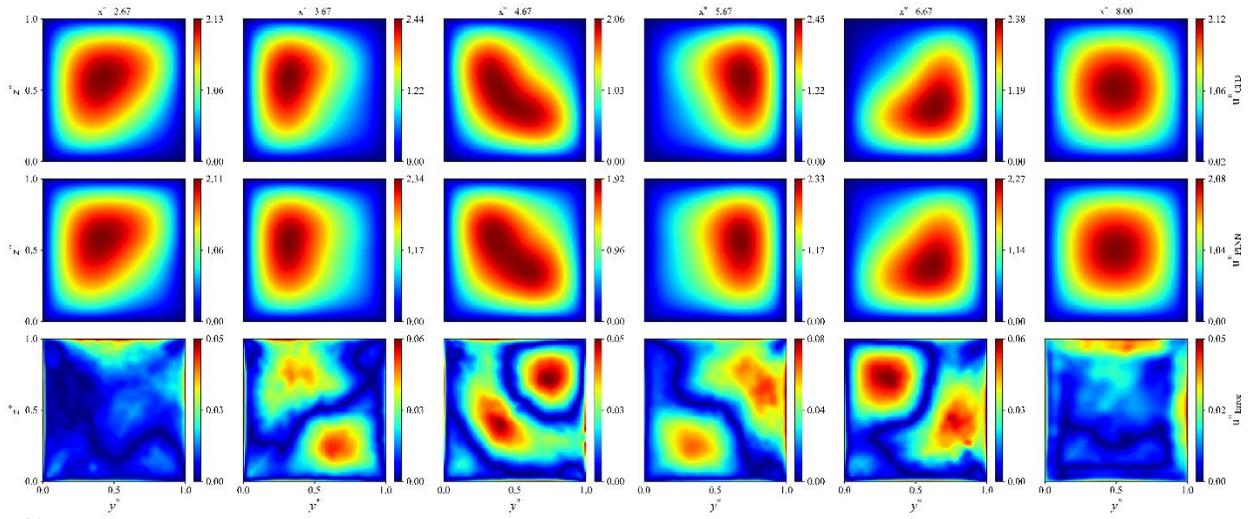

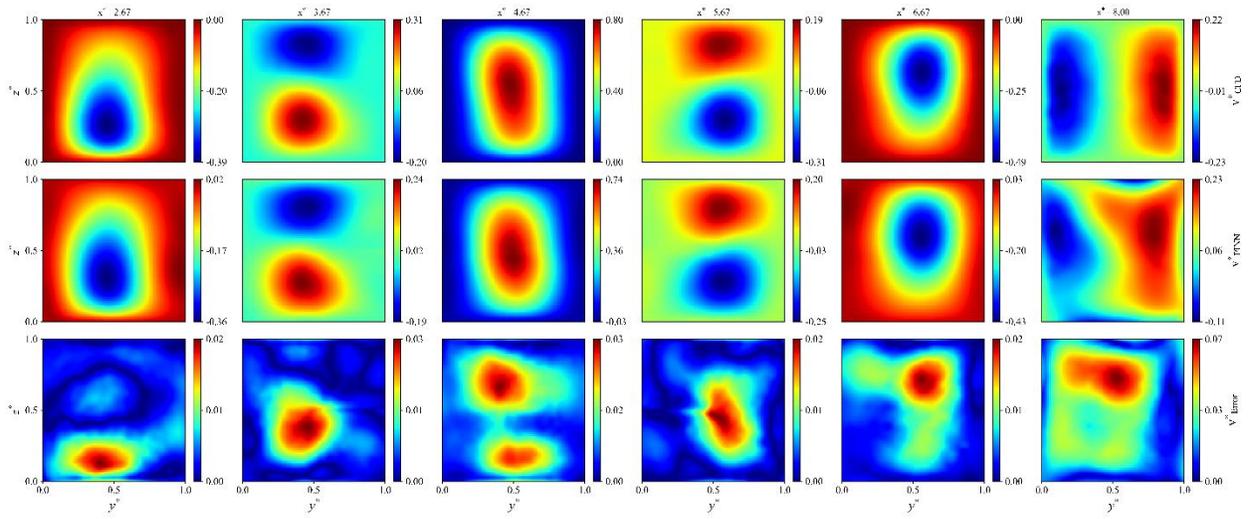

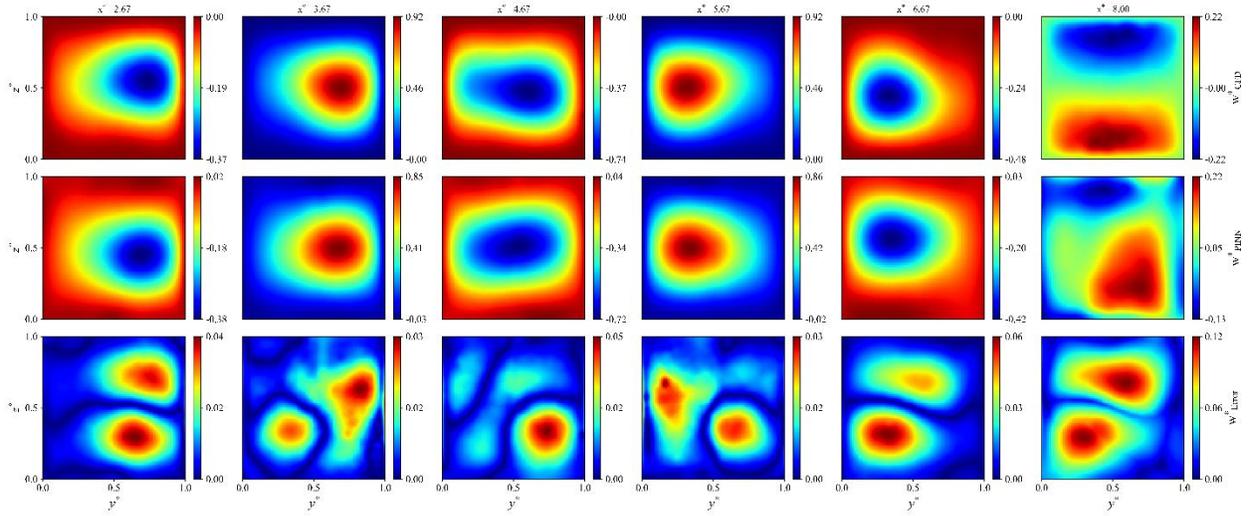

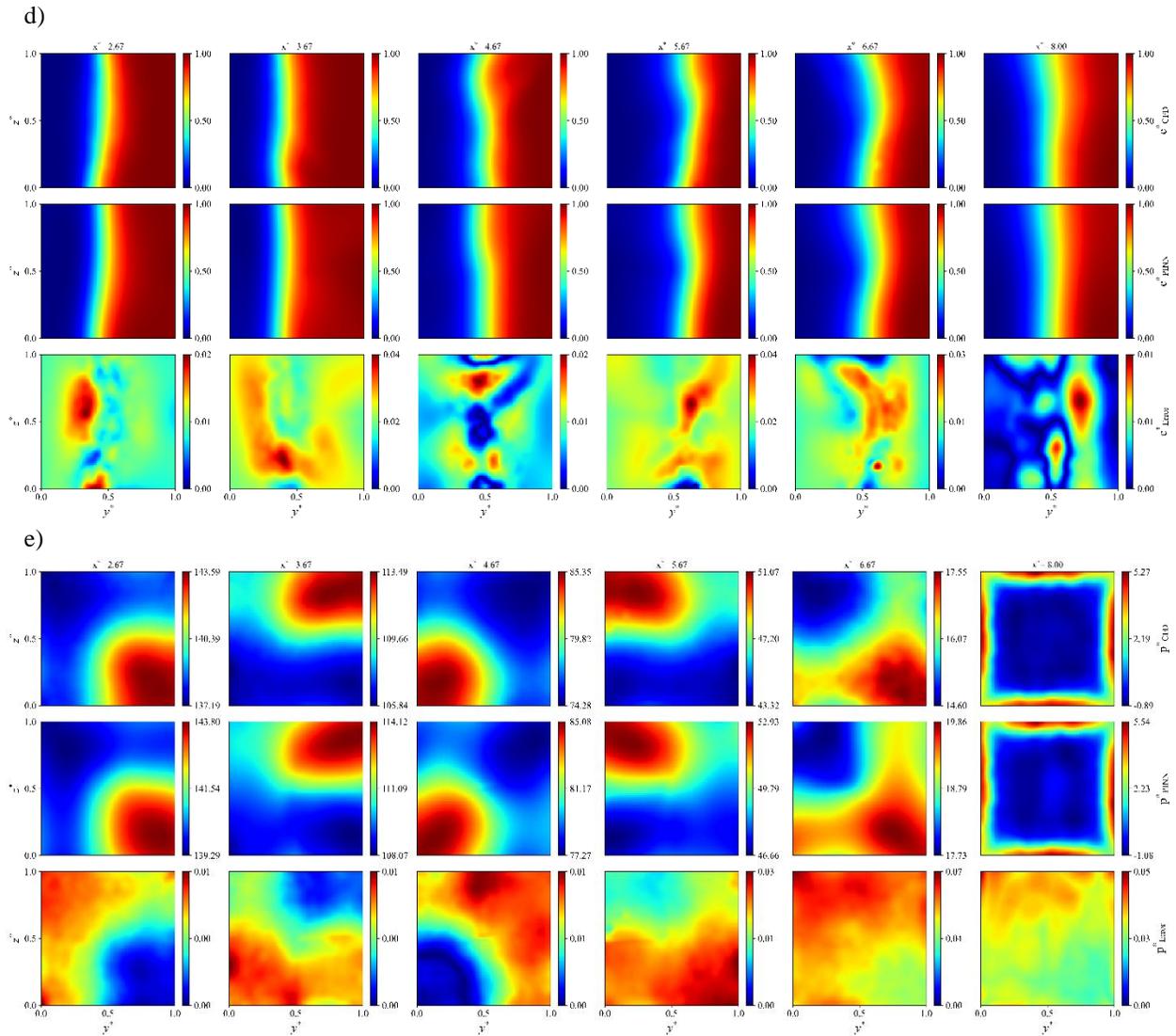

Figure 8: Comparison of results obtained from the FlexPINN method with CFD results at various cross-sections of the channel in Configuration A. a) Dimensionless horizontal velocity, b) Dimensionless vertical velocity, c) Dimensionless in-plane velocity, d) Dimensionless concentration, e) Dimensionless pressure.

### Effect of Fin Configurations

The effect of different arrangements is examined in Figure 9, where the contour of dimensionless concentration for various configurations in a one-unit channel with rectangular fins is analyzed. These contours are displayed at the middle section of the channel in both the xy and xz planes. Contour plots for all cases, including dimensionless velocity and dimensionless pressure components for different configurations, are provided on the [GitHub repository](#).

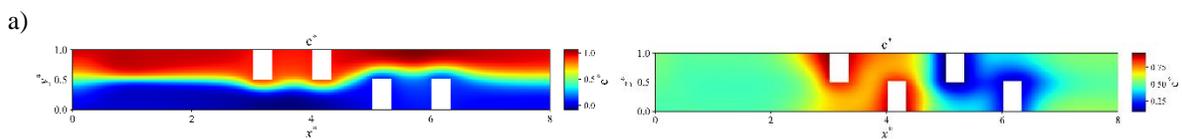

a)

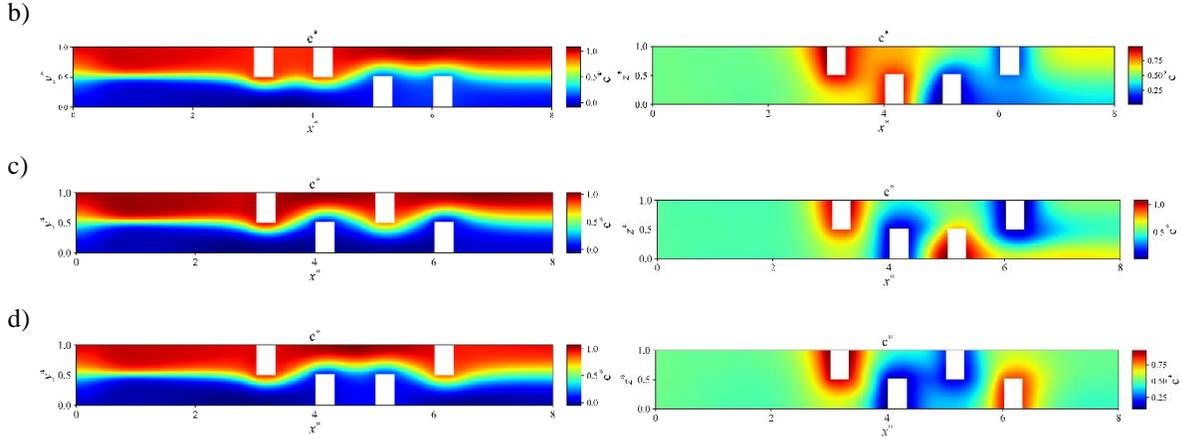

Figure 9: Contours of dimensionless concentration for fin configurations in the xy-plane (left) and xz-plane (right). a) Configuration A, b) Configuration B, c) Configuration C, d) Configuration D.

To comprehensively assess the influence of fin arrangement on micromixing performance, Figure 10 illustrates the effect of various fin configurations and Reynolds numbers on the pressure drop coefficient, mixing index, and mixing efficiency (as defined by Equation 38) [20]. In this equation, the subscript zero denotes the reference case of a finless channel, which was used for initial validation purposes. As depicted in the figure, increasing the Reynolds number results in a general decline in the pressure drop coefficient across all configurations, which can be attributed to the relatively thinner boundary layers and more streamlined flow paths at higher flow rates. Among the configurations, Configuration B yields the lowest pressure drop. This outcome is likely due to its more orderly and symmetric fin layout, which minimizes flow separation and suppresses the generation of strong secondary flows, thereby reducing pressure loss.

In contrast, Configuration C achieves the highest mixing index—an outcome that aligns with the primary objective of the micromixer. The irregular, staggered placement of fins along the y-axis in this configuration promotes chaotic advection and enhances fluid interfacial stretching, both of which are critical mechanisms for effective mixing in laminar regimes. Given that the two inlet streams are introduced along this same axis, the alternating fin placement significantly increases the contact surface and interaction time between the fluids, thereby accelerating homogenization. To provide a balanced assessment of mixing quality versus energy cost, the mixing efficiency parameter introduced in Equation 38 relates the improvement in mixing to the associated pressure penalty. When this parameter exceeds unity, it suggests that the enhancement in mixing justifies the incurred pressure loss, indicating a practical and efficient fin design. Configuration C not only demonstrates superior mixing performance but also achieves the highest mixing efficiency among the examined setups. This makes it the most promising candidate for optimized micromixer design and motivates its selection for detailed analysis in the subsequent sections of this study.

$$ME = \frac{\frac{MI}{MI_0}}{\left(\frac{C_p}{C_{p,0}}\right)^{\frac{1}{3}}} \tag{38}$$

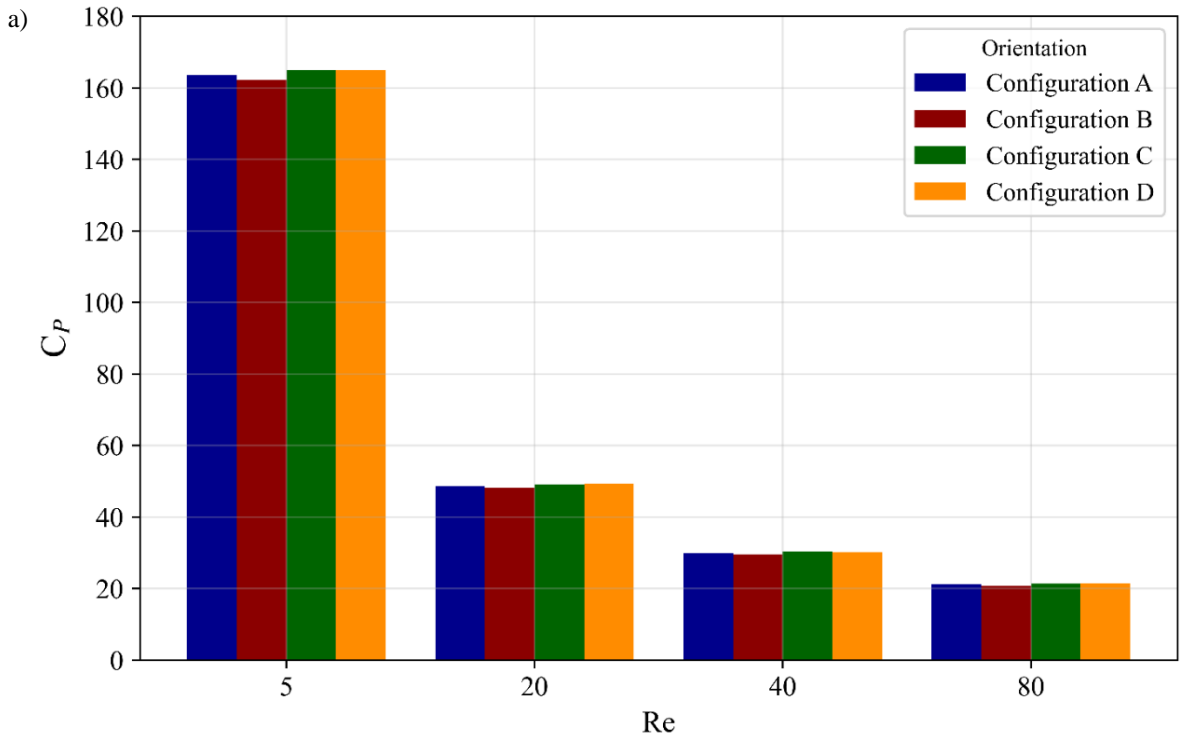

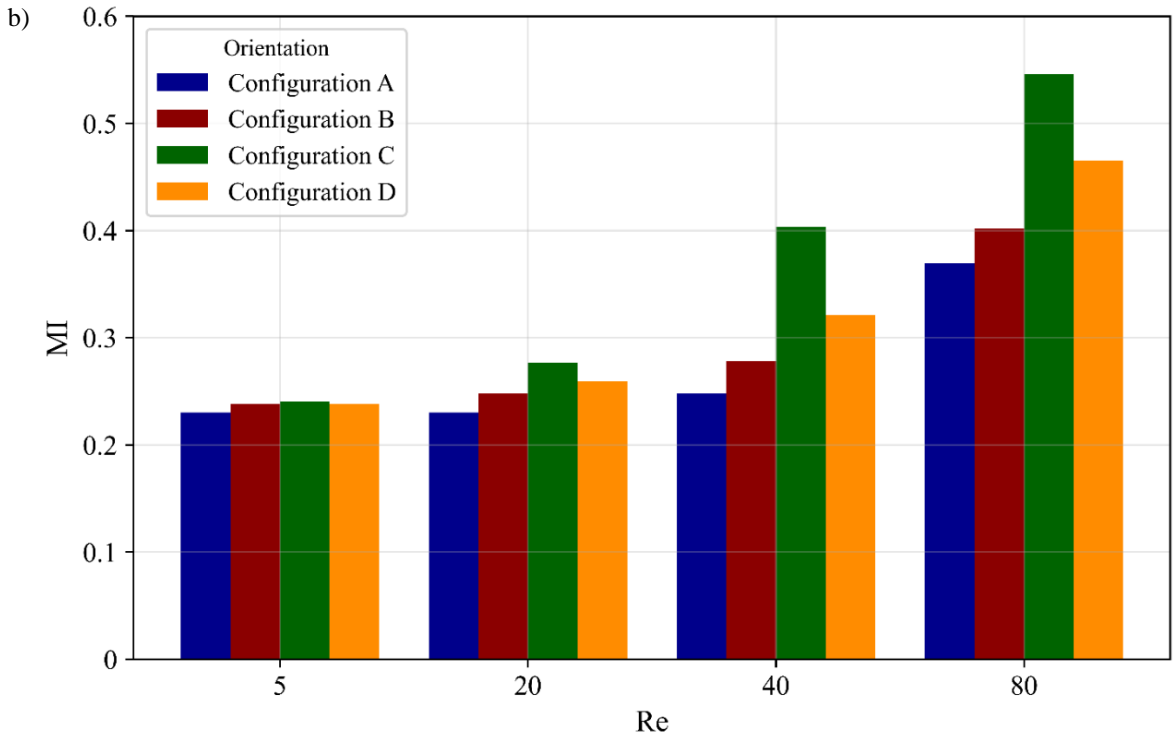

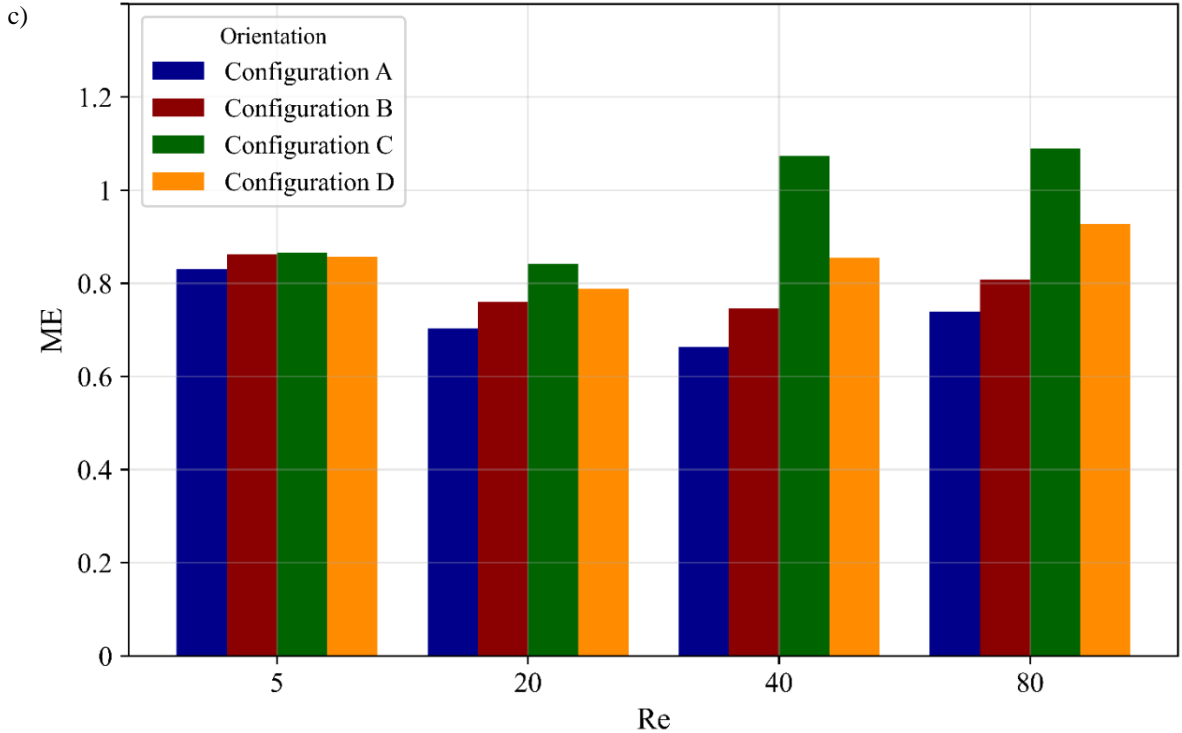

Figure 10: Effect of fin configuration and Reynolds number on: a) Pressure drop coefficient, b) Mixing index, c) Mixing efficiency.

## Effects of Fin Shape in a Single-Unit Channel

Following the previous analysis, Configuration C demonstrated the highest mixing efficiency among all fin configurations. Consequently, this configuration was selected for further evaluation involving different fin shapes. Figure 11 presents the impact of fin geometry and Reynolds number on the pressure drop coefficient, mixing index, and mixing efficiency within a single-unit channel. Across all three fin shapes, an increase in Reynolds number generally leads to a decrease in pressure drop coefficient and a simultaneous increase in the mixing index. According to the classification by Rasouli et al. [15], micromixing based on Reynolds number can be divided into three distinct regimes: diffusive, transient, and chaotic. The Reynolds numbers used in this study span all three regions, and the transition between these regimes is clearly observable in the plotted results. As Reynolds number increases, the overall mixing efficiency tends to improve. However, a notable exception is observed in the case of rectangular fins, where an initial decrease followed by an increase in mixing efficiency occurs at lower Reynolds numbers. This non-monotonic behavior is attributed to the dominance of diffusive mixing mechanisms at low Reynolds numbers, where convective effects are minimal. In this regime, the pressure penalty introduced by rectangular fins outweighs the gains in mixing, leading to a temporary drop in efficiency. In contrast, the elliptical and triangular fins—due to their streamlined profiles—exert a lower pressure drop and thus show a steadier increase in mixing efficiency as the Reynolds number rises from 5 to 20. Nevertheless, the improvement is relatively modest in these two cases due to the limited enhancement in interfacial stretching under low-inertia flow.

Among the three geometries, the rectangular fins consistently produce the highest mixing index across all Reynolds numbers. This is likely due to their sharp edges and bluff-body nature, which promote stronger local recirculation zones and more intense stretching and folding of fluid interfaces—key mechanisms for enhancing scalar transport in laminar flows. While this comes at the cost of a higher pressure drop, the net result is a higher mixing efficiency. Furthermore, in the lower Reynolds number range (below Re = 30), corresponding to the diffusive and early transient regimes, elliptical fins outperform triangular ones in terms of mixing performance. This can be attributed to the smoother, more continuous curvature of elliptical fins, which enhances lateral flow development and promotes more uniform stretching of concentration gradients. In contrast, the sharper angles of triangular fins are more effective at higher Reynolds numbers, where inertial forces dominate. Their geometry promotes the formation of strong, localized vortices and flow separation, which enhance mixing through chaotic advection mechanisms, making them particularly suitable for chaotic mixing regimes beyond the diffusive range.

In summary, although rectangular fins induce the highest pressure loss, they also achieve the best mixing due to their capacity to disrupt flow and generate strong internal vortices. Elliptical fins, strike a better balance in the diffusive regime by gently enhancing mixing while maintaining a lower pressure penalty, making them preferable for applications operating at low Reynolds numbers. Triangular fins on the other hand, perform better in the chaotic regime because their angled structure intensifies vortex shedding and streamwise flow disruption, which are key drivers of mixing under high-Re conditions.

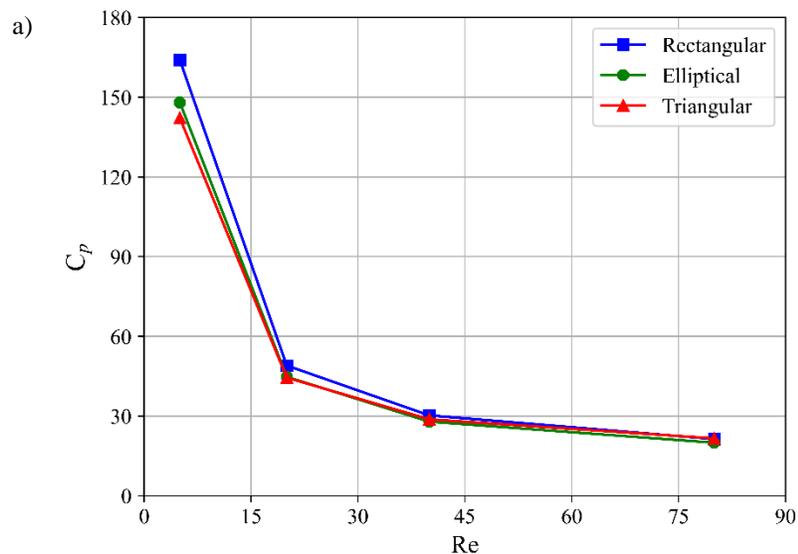

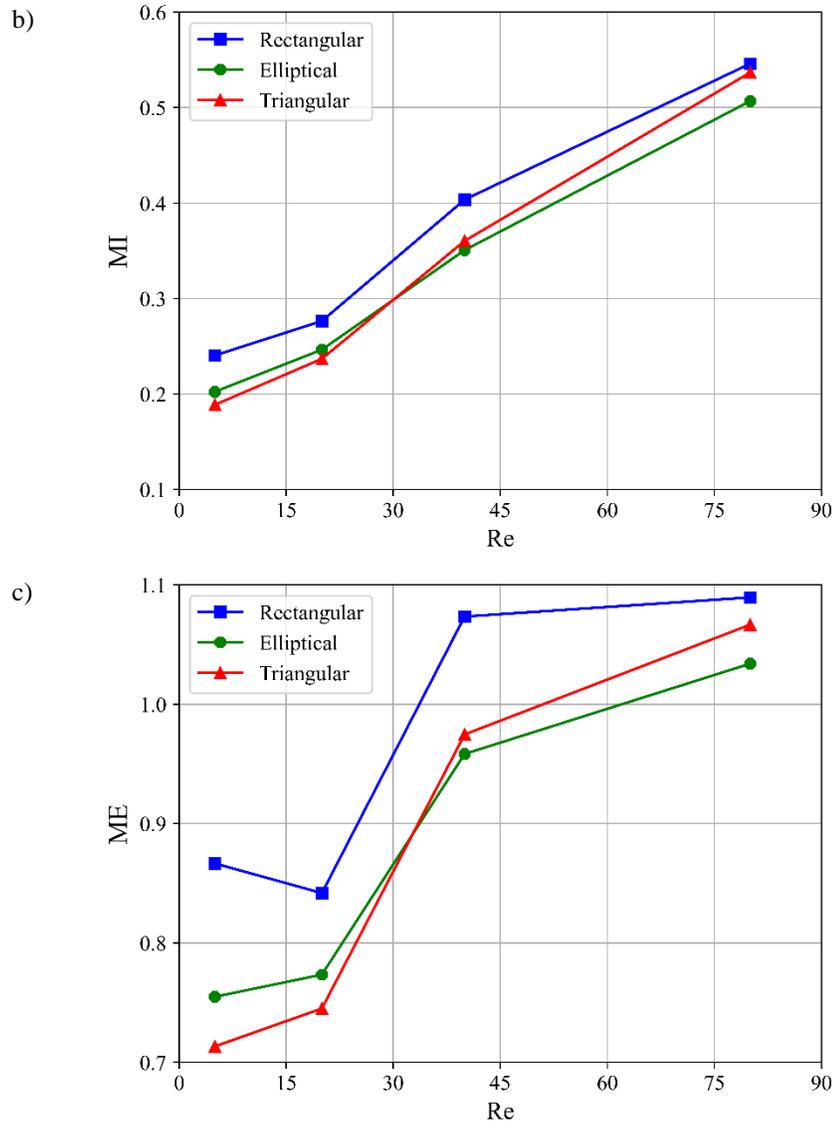

Figure 11: Influence of Reynolds number and fin shape on: (a) Pressure drop coefficient, (b) Mixing index at the channel outlet, and (c) Mixing efficiency.

## Effects of Fin Shape in a Double-Unit Channel

In the final part of this study, the double-unit channel is also evaluated. Figure 12 illustrates the distribution of dimensionless concentration in the xy and xz planes for three different fin shapes.

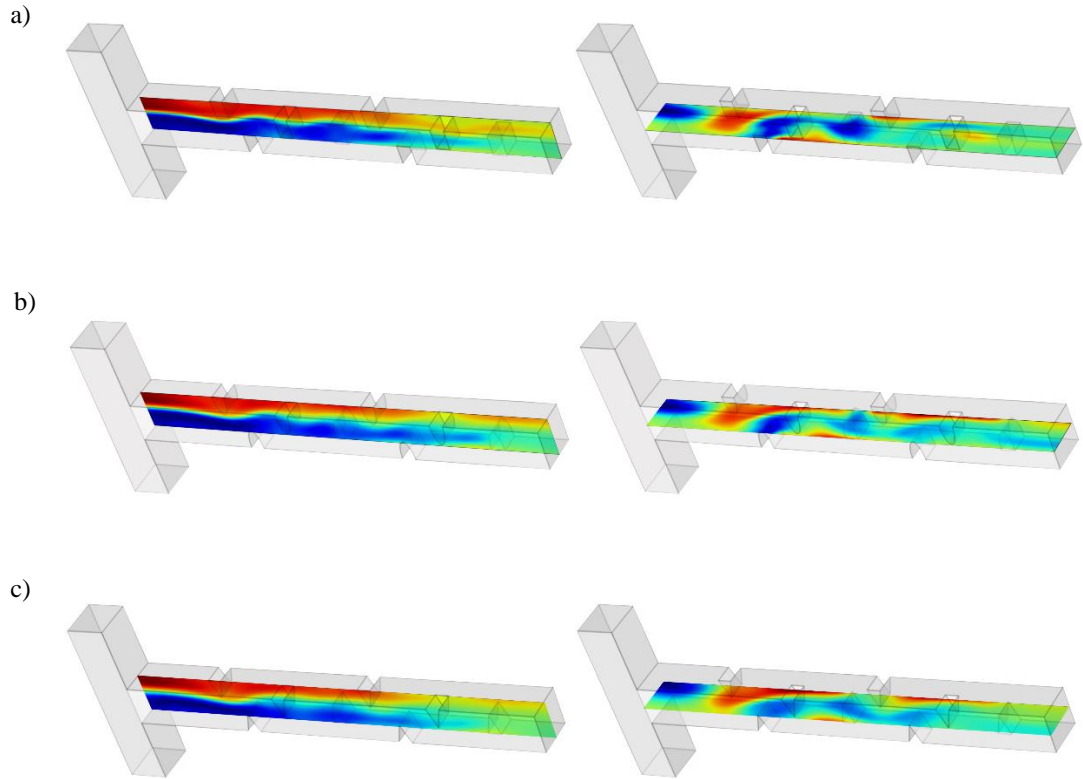

Figure 12: Dimensionless concentration contours in the xy-plane (left) and xz-plane (right): (a) Rectangular fins, (b) Elliptical fins, (c) Triangular fins.

Figure 13 presents the dimensionless concentration contours along the channel at Reynolds number 80 for all three fin shapes. Additional results, including dimensionless velocity and pressure contours along the channel, are available at the GitHub link.

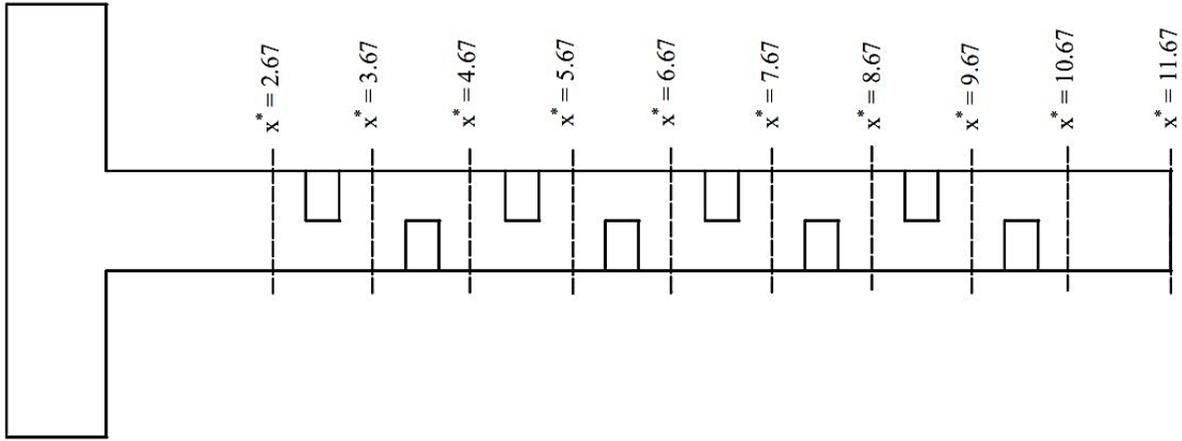

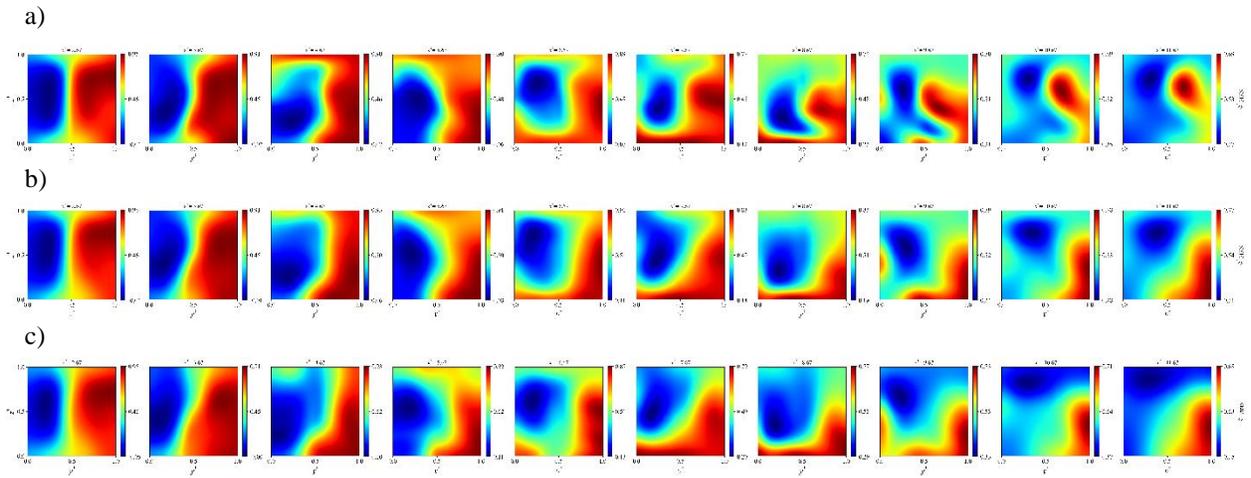

Figure 13: Dimensionless concentration contours along the channel at Reynolds number 80: (a) Rectangular fins, (b) Elliptical fins, (c) Triangular fins.

The effect of Reynolds number on the mixing index along the channel is shown in Figure 14 for all three fin shapes. As seen in the figure, increasing the Reynolds number leads to a higher mixing index throughout the channel in all fin configurations. Reynolds number 5 corresponds to the diffusive regime, while Reynolds number 80 falls within the chaotic regime. In the transitional regime, represented by Reynolds numbers 20 and 40—which mark the beginning and end of this region in the current study—a considerable fluctuation in the mixing index is observed. This variation highlights that the most rapid changes in mixing performance with respect to Reynolds number occur in the transitional regime, where the flow starts to shift from being diffusion-dominated to inertia-driven. Such behavior emphasizes the critical role of this region in micromixer operation, where even small changes in Reynolds number can significantly impact mixing effectiveness.

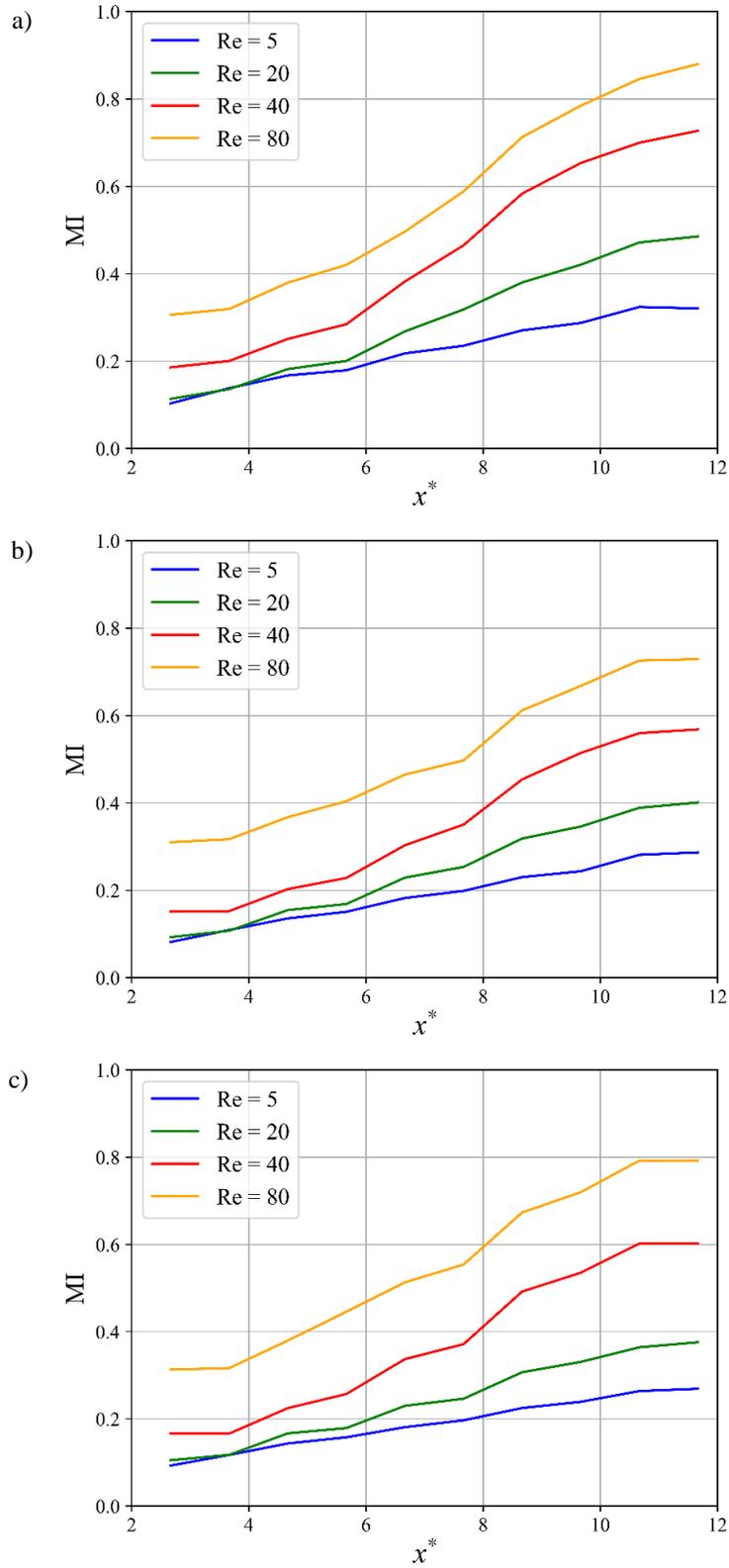

Figure 14: Effect of Reynolds number on the mixing index along the channel for: a) Rectangular fins, b) Elliptical fins, and c) Triangular fins.

The effect of Reynolds number and fin shape on pressure drop coefficient, mixing index, and mixing efficiency in the two-unit channel is illustrated in Figure 15. As the Reynolds number increases, mixing index rise for all three fin geometries, while pressure drop coefficient decreases. Beyond Reynolds numbers of 40, the mixing efficiency shows a declining trend. Notably, elliptical fins outperform triangular fins in terms of mixing efficiency at Reynolds numbers below 30, while rectangular fins consistently achieve the highest mixing efficiency across all tested Reynolds numbers. Due to the extended length of the two-unit channel, when the Reynolds number exceeds 40, the increasing pressure losses along the channel begin to outweigh the gains in mixing index, leading to a reduction in overall performance across all fin types. While higher Reynolds numbers enhance the mixing of fluids within the micromixer, the pressure drop increases at a faster rate than the improvement in mixing index, thereby reducing the overall efficiency.

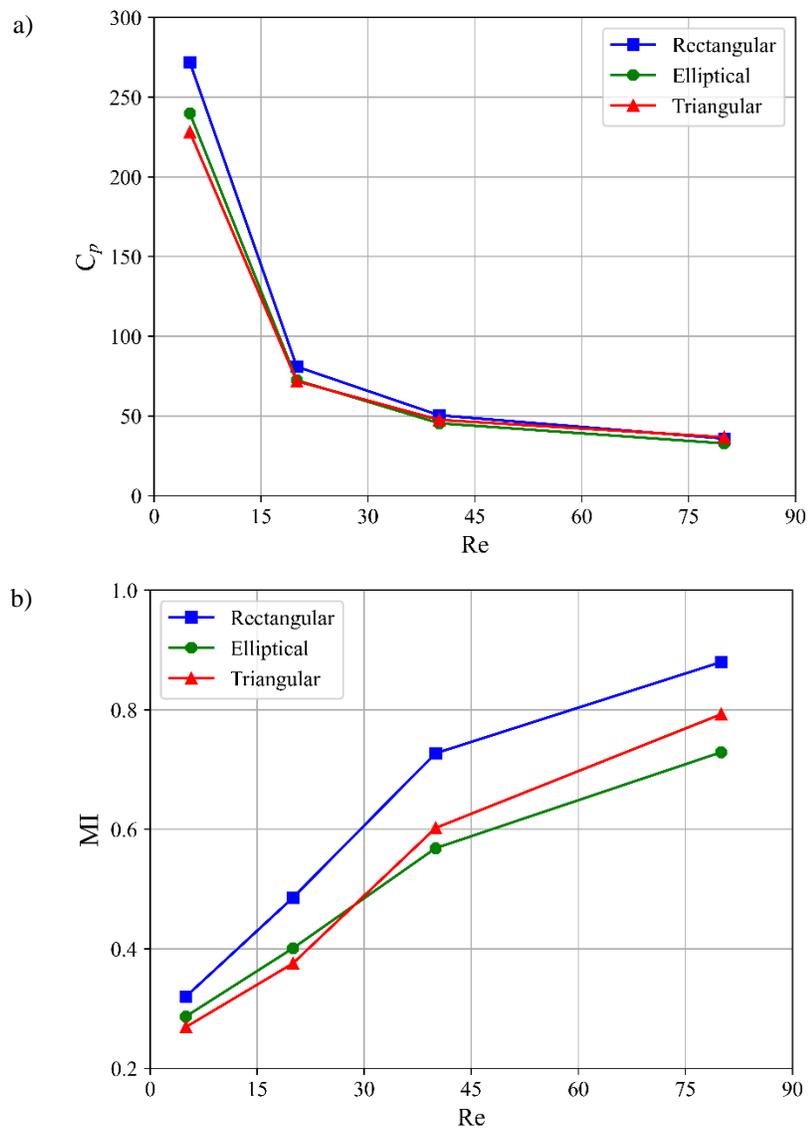

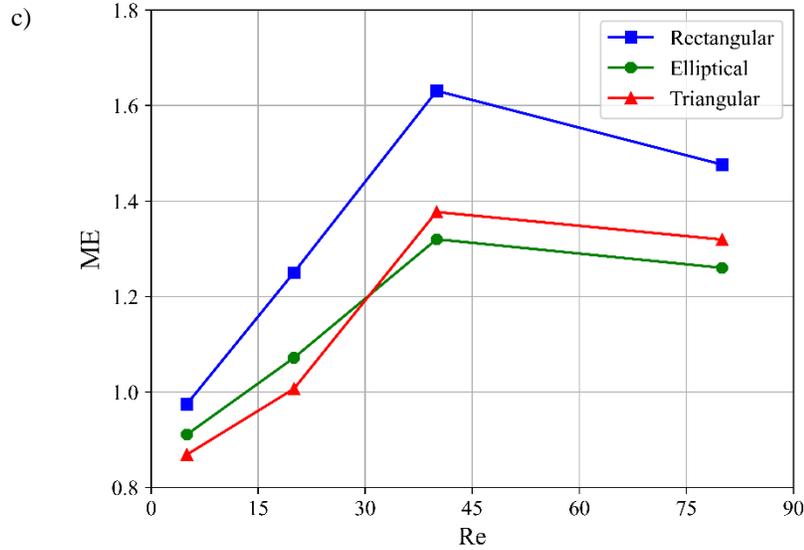

Figure 15: Effect of Reynolds number and fin geometry in the two-unit channel on: a) Pressure drop coefficient, b) Mixing index at outlet of channel, and c) Mixing efficiency.

In this study, the FlexPINN framework has successfully simulated fluid flow and mass transfer phenomena in a 3D microchannel equipped with fins. This model can be further extended through parametric studies to explore the full potential of the method. Moreover, incorporating data from other numerical approaches for use in transfer learning [63], or integrating with alternative neural network structures [64], may offer valuable avenues for enhancing the capabilities of FlexPINN.

## Conclusion

Physics-Informed Neural Networks (PINNs) have emerged as an advanced method in numerical computation, demonstrating significant potential in solving complex physical problems governed by partial differential equations. By embedding governing physical laws directly into the training process, PINNs eliminate the need for mesh generation, offering a flexible, mesh-free, and generalizable framework. Traditional numerical methods often require extensive meshing and substantial computational resources, particularly in three-dimensional domains with complex internal geometries. PINNs, by leveraging automatic differentiation and physical constraints, can efficiently simulate fluid dynamics without relying on large datasets or discretization schemes—making them an ideal candidate for such problems. The application of PINNs in microfluidic systems is especially compelling, as these systems exhibit intricate flow behaviors, steep concentration gradients, and pressure variations at microscale levels.

In this study, an enhanced version of PINN, named FlexPINN, was effectively employed to simulate incompressible 3D fluid flow and mass transfer in microchannels with various fin geometries. FlexPINN demonstrated high accuracy in predicting the distribution of concentration, pressure, and velocity, with results aligning closely with expected physical trends. Moreover, it showed significantly better convergence behavior than the vanilla PINN, delivering solutions consistent with the underlying physics. From a computational efficiency perspective, the use of transfer learning within FlexPINN led to up to 35% reduction in solution time for elliptical and triangular fins at low Reynolds numbers compared to rectangular fins, reinforcing the method's capability for high-precision large-scale simulations.

The outcomes of this study provide valuable insights into the relationship between fin geometry and mixing performance as well as pressure drop in microchannels. Among the tested fin shapes— rectangular, elliptical, and triangular— rectangular fins consistently exhibited the highest mixing index across all Reynolds numbers, while elliptical fins outperformed triangular ones at lower Reynolds regimes. This research also highlights the critical role of fin arrangement in optimizing flow characteristics. The so-called type-C configuration delivered the best overall performance, achieving the highest mixing efficiency. This configuration, characterized by a more irregular fin layout along the channel, enhanced fluid mixing due to improved disruption of laminar layers. Analysis of key performance metrics such as mixing index and pressure drop coefficient reveals that geometric variations in microchannel design can significantly influence system efficiency— an aspect of great importance for optimizing devices in biomedical, chemical, and energy-related applications. Ultimately, this study underscores the strong flexibility of FlexPINN in solving multiphysics problems involving simultaneous momentum and mass transfer. Its ability to accurately model complex physical phenomena, such as microscale mixing and pressure variations, positions it as a powerful tool for the future design of microfluidic systems. This novel approach has the potential to revolutionize the design, testing, and optimization of microfluidic devices, opening new directions for both research and real-world applications in fields like biomedical diagnostics and environmental monitoring.

### CRediT authorship contribution statement
Meraj Hassanzadeh: Conceptualization, Methodology, Software, Validation, Investigation, Visualization, Writing the original draft.
Ehsan Ghaderi: Conceptualization, Methodology, Software, Validation, Investigation, Visualization, Writing the original draft.
Bijarchi et al.: Conceptualization, Methodology, Review and Editing, Supervision.

### Acknowledgment
The authors express their gratitude to the Deputy of Research and Technology of Sharif University for providing a suitable working environment to conduct the experiments and acknowledge Miss. Tina Hajihadi Naghash & Mr. Amir Mohammad Haghgoo for their helpful and kind support.

### Conflict of Interest
The authors declare no conflict of interest.

### Data Availability
The python codes used to generate the results presented are available on Github.
https://github.com/imRaajee/FlexPINN